\documentclass[
reprint,
aps,
pra,
superscriptaddress,
amsmath,
amssymb,
nobalancelastpage,
floatfix
]{revtex4-2}

\usepackage{graphicx}
\usepackage{dcolumn}
\usepackage{bm}

\usepackage[utf8]{inputenc}
\usepackage[T1]{fontenc}

\usepackage[english]{babel}
\usepackage[left=0.8in,right=0.8in,top=0.8in,bottom=0.8in]{geometry}
\usepackage{amssymb}
\usepackage{amsmath}
\usepackage[version=4]{mhchem}
\usepackage{bbold}
\usepackage{dsfont}
\usepackage{mathtools}
\usepackage{amsthm}
\usepackage{subcaption}
\usepackage{multirow}
\usepackage{booktabs}  
\usepackage{xcolor}
\usepackage{bm}
\usepackage{braket}
\usepackage{tikz}
\usetikzlibrary{quantikz}
\usepackage{physics}
\usepackage[hidelinks]{hyperref}
\usepackage{cleveref}
\let\emph\relax
\DeclareTextFontCommand{\emph}{\bfseries}

\definecolor{blue}{HTML}{692e9e}
\definecolor{red}{HTML}{d94545}

\makeatletter
\def\@email#1#2{%
 \endgroup
 \patchcmd{\titleblock@produce}
  {\frontmatter@RRAPformat}
  {\frontmatter@RRAPformat{\produce@RRAP{*#1\href{mailto:#2}{#2}}}\frontmatter@RRAPformat}
  {}{}
}%
\makeatother

\begin{document}

\preprint{AIP/123-QED}

\title{Folded Spectrum VQE : A quantum computing method for the calculation of molecular excited states}

\author{Lila Cadi Tazi}
\affiliation{Yusuf Hamied Department of Chemistry, University of Cambridge, Cambridge, UK}
\affiliation{École Normale Supérieure Paris-Saclay, Université Paris-Saclay, Gif-sur-Yvette, France}

\author{Alex J.W. Thom}
\affiliation{Yusuf Hamied Department of Chemistry, University of Cambridge, Cambridge, UK}

    \begin{abstract}
      The recent developments of quantum computing present novel potential pathways for quantum chemistry, as the scaling of computational power of quantum computers could be harnessed to naturally encode and solve electronic structure problems. Theoretically exact quantum algorithms for chemistry have been proposed (e.g. Quantum Phase Estimation), but the limited capabilities of current noisy intermediate-scale quantum devices (NISQ) motivated the development of less demanding hybrid algorithms. In this context, the Variational Quantum Eigensolver (VQE) algorithm was successfully introduced as an effective method to compute the ground-state energy of small molecules. This study investigates the Folded Spectrum (FS) method as an extension to the VQE algorithm for the computation of molecular excited states. It provides the possibility of directly computing excited states around a selected target energy using the same ansatz as for the ground-state calculation. Inspired by the variance-based methods from the Quantum Monte Carlo literature, the FS method minimizes the energy variance, thus, in principle, requiring a computationally expensive squared Hamiltonian to be applied. We alleviate this potentially poor scaling by employing a Pauli grouping procedure, identifying sets of commuting Pauli strings that can be evaluated simultaneously. This allows for a significant reduction in computational cost. We applied the FS-VQE method to small molecules ($\ce{H2}$, $\ce{LiH}$), obtaining all electronic excited states with chemical accuracy on ideal quantum simulators. Furthermore, we explore the application of quantum error mitigation techniques, demonstrating improved energy accuracy on noisy simulators compared to simulations without mitigation. 
    \end{abstract}

    \maketitle
    
\begin{figure*}[ht]
    \centering
    \includegraphics[width=0.8\linewidth]{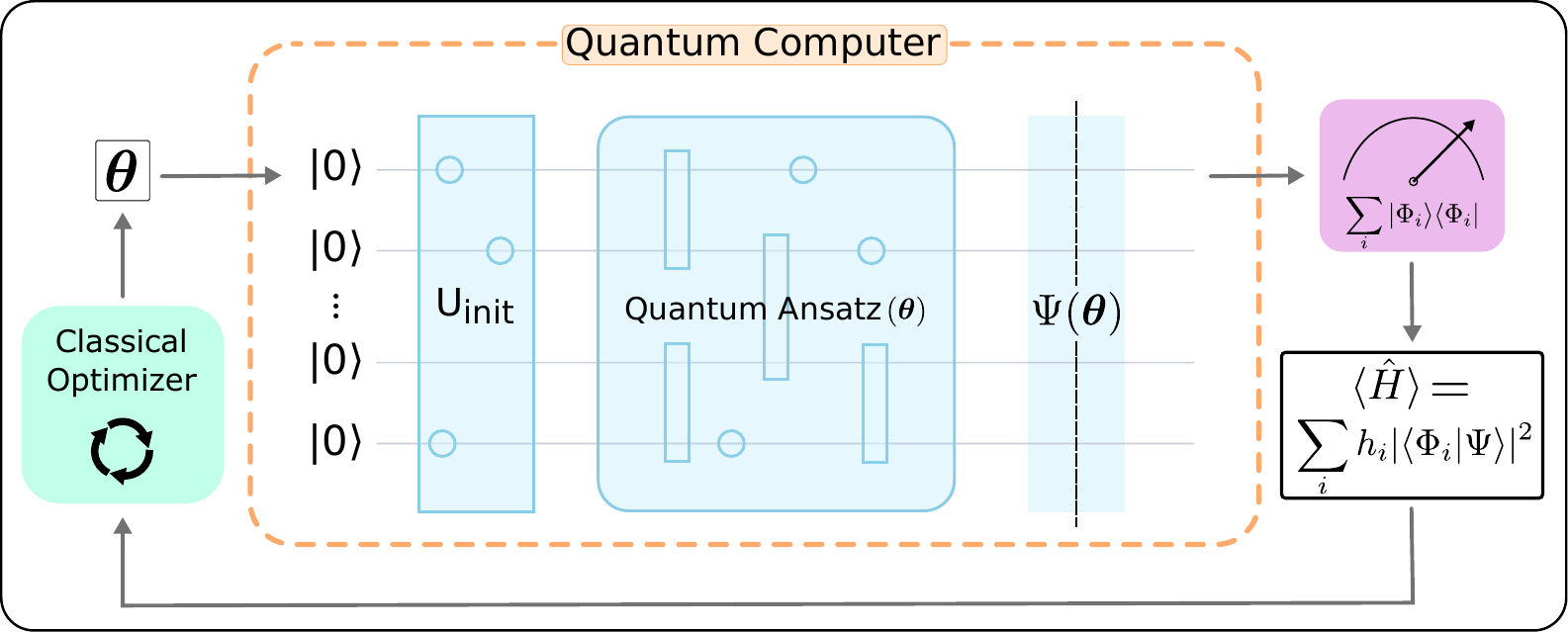}
    \caption{Principle of VQE algorithm}
    \label{fig:vqe}
\end{figure*}

\section{Introduction}
Computing the electronic structure of molecules and materials is crucial for the prediction of chemical or structural properties. Theoretical chemists and physicists have acknowledged the essential challenges that must be addressed, but the exponentially scaling dimensionality of electronic structure problems seems to be insurmountable on classical computing resources. As a result, the theoretical study of large molecules or materials using \textit{ab initio} methods such as coupled cluster is impractical. Hence, less costly methods involving approximations are generally employed at the cost of a loss in accuracy and predictive power. The emergence of quantum computing presents potential novel pathways for theoretical chemistry, as quantum resources show exponentially scaling computational power that could be harnessed to naturally encode and solve quantum problems. Although exponential speedup may not be achieved, a polynomial acceleration could be ground-breaking for quantum chemistry applications \cite{Lee2023}.

In this context, the Variational Quantum Eigensolver (VQE) \cite{Peruzzo2014} was introduced as an effective algorithm to find the lowest eigenvalue of a quantum observable. In particular, it can compute the ground-state energy of a molecular Hamiltonian. The capability of VQE for electronic ground-state computation of small molecules has been extensively studied \cite{VQEReview2021}, but the effective and direct computation of excited states remains elusive.

In this study, we propose a variant of VQE that aims at computing molecular excited states. It uses the Folded Spectrum (FS) method to reorder the Hamiltonian's eigenspectrum, thus allowing for the direct computation of highly excited states. Although this method is documented in the literature, its quantum implementation was considered too costly due to the squared number of terms of the measured operator \cite{Santagati2018, Zhang2021}. Here, we show that a Pauli grouping procedure reduces the required number of measurements, thereby making the cost of the FS method reasonable. The effect of Pauli grouping is particularly significant for second quantized molecular Hamiltonians as a result of their particular structure. Finally, we present FS-VQE results obtained on a noisy quantum simulator, and show the successful use of quantum error mitigation techniques on this algorithm. 

\section{Variational Quantum Eigensolver}

The Variational Quantum Eigensolver \cite{Peruzzo2014} is a hybrid quantum-classical algorithm (see figure \ref{fig:vqe}). Its purpose is to find the lowest eigenvalue of a given quantum operator. It can be applied to quantum chemistry problems to obtain the electronic ground-state of a molecule, by focusing on the molecular Hamiltonian $\hat{H}$.

The algorithm relies on an \emph{ansatz} to prepare a trial electronic wavefunction on the quantum computer. The ansatz takes the form of a parametric quantum circuit whose parameters, denoted $\bm\theta$, are angles in rotation gates. Details on the ansatz design are given in section \ref{sec:ansatz}.

The quantum subroutine prepares a parametric trial wavefunction $\Psi(\bm\theta)$ on a qubit register. This quantum state can be assessed by measuring the qubits : from the measurement results, the expectation value of the molecular Hamiltonian $\langle \hat{H} \rangle$ can be computed on a classical computer (see section \ref{subsec:expvalcomp}). This value corresponds to the average electronic energy of the trial wavefunction $\Psi(\bm\theta)$.
A classical optimizer is then used to adjust the parameters $\bm\theta$ in the ansatz in order to minimise the value of $\langle \hat{H} \rangle$. By means of the variational principle, the minimal expectation value obtained for a set of parameters $\bm\theta_{\mathrm{opt}}$ is an upper bound on the Hamiltonian's ground energy. The quantum state prepared with the optimal, final angles $\bm\theta_{\mathrm{opt}}$ is a representation of the molecule's ground-state electronic wavefunction.

\subsection{Ansatz}\label{sec:ansatz}

In the context of VQE, the ansatz is a parametric quantum circuit aiming to explore the wavefunction search space. The ansatz design can take various forms as different properties are targeted \cite{Cao2019}. 
The number of parameters in the circuit is key to the success of the optimization procedure ; a very large number of parameters may lead to intractable optimization.

The so called \emph{chemically motivated ansatz} class includes ansätze inspired by quantum chemistry methods \cite{Cao2019}. Their advantage is that the prepared states are by design physically relevant (number of electrons and total spin are conserved).
However, they often require a large number of parameters and deep quantum circuits, which limits both the optimization success and their feasibility on NISQ hardware.

Another approach is to design \emph{hardware motivated ansätze} \cite{Cao2019}. Such ansätze are constructed to be efficiently implemented on quantum computers. 
Strong constraints in terms of quantum gates, qubit connectivity, number of two-qubit gates, global circuit depth, etc. are defined in accordance with the capability of the target hardware. 
These ansätze are computationally advantageous, but they do not offer a guarantee on the physical properties of the prepared trial states, thus limiting the convergence.

\subsubsection{Unitary Coupled Cluster Ansatz}
Unitary coupled cluster (UCC) is a widely used chemically motivated ansatz for electronic wavefunctions in quantum computing. 
It is a unitary variant of the well-known coupled cluster (CC) theory. 
Like coupled cluster, UCC is based on a reference wavefunction (often Hartree--Fock) and it creates linear combinations of excited determinants using excitation operators $\hat{T}$:
\begin{equation}
    \hat{T} = \hat{T}_1 + \hat{T}_2 + \hat{T}_3 + \hat{T}_4 +...
\end{equation}

\begin{equation}\label{eq:t1}
    \hat{T}_1 = \sum_{i;a} \theta_i^a \;\hat{a}^\dagger_a \hat{a}_i
\end{equation}

\begin{equation}\label{eq:t2}
    \hat{T}_2 = \sum_{i<j;a<b} \theta_{ij}^{ab} \; \hat{a}^\dagger_a\hat{a}^\dagger_b \hat{a}_i\hat{a}_j
\end{equation}

\noindent where $\hat{T}_1$ is the operator of all single excitations, $\hat{T}_2$ the operator of all double excitations, etc. $\hat{a}^\dagger_k$ and $\hat{a}_k$ are, respectively, the fermionic creation and annihilation operators acting on orbital $k$. Indices $i,j$ denote occupied orbitals and $a,b$ virtual orbitals.
Parameters $\bm\theta$ are optimized to obtain the CC wavefunction.

Because the CC operator $e^{\hat{T}}$ is not unitary, it cannot be directly implemented on a quantum circuit.
To create a unitary variant of CC, the cluster operator needs to be modified to become anti-hermitian, as the exponentiation of an anti-hermitian operator is unitary:

\begin{equation}
    \hat{O}^\dagger = - \hat{O} \; \Rightarrow \; e^{\hat{O}}{e^{\hat{O}}}^\dagger = \mathbb{1}.
\end{equation}

Therefore, the anti-Hermitian cluster operator $\hat{T}-\hat{T}^\dagger$ is considered. The UCC ansatz state is created similarly to the CC state :
\begin{equation}
    \ket{\Psi_{\mathrm{UCC}}} = e^{\hat{T}-\hat{T}^\dagger} \ket{\Psi_0}.
\end{equation}

For the UCC state to serve as a quantum ansatz, the operator $e^{\hat{T}-\hat{T}^\dagger}$ must be expressed in terms of quantum gates. Since excitation operators do not commute, a Trotterization step is required to decompose the exponentiated operator \cite{UCC2022}. The Trotter decomposition is given by :

\begin{equation}
\resizebox{0.9\columnwidth}{!}{
    $e^{\hat{T}-\hat{T}^\dagger}=e^{\sum_i \theta_i (\hat{T}_i-\hat{T}_i^\dagger)} = \left( \prod_i e^{\frac{\theta_i}{p} (\hat{T}_i-\hat{T}_i^\dagger)}\right)^p+\mathcal{O}\left(\frac{1}{p}\right)$}
\end{equation}

\noindent with $\hat{T}_i$ the excitation operators defined in eq. (\ref{eq:t1}) and (\ref{eq:t2}) and $p$ the Trotter decomposition order. In this work, the order $p=1$ was implemented, as it was proven to be an exact and general form of UCC under the condition of an appropriate ordering of excitations \cite{Evangelista2019} (details in section \ref{uccorder}).
This large form of UCC is truncated to a rank $k$ corresponding to the highest excitation considered (for example, UCCSD is $k = 2$ : only single and double excitations). 
The resulting smaller ansatz is, therefore, : 

\begin{equation}
     \ket{\Psi_{\mathrm{UCC}}} \approx \prod_{i=1}^k e^{ \theta_i (\hat{T}_i-\hat{T}_i^\dagger)}  \ket{\Psi_0}
\end{equation}

\noindent with parameters $\bm\theta$ to optimize. 
\smallskip

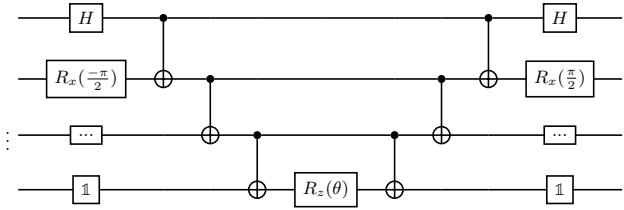
\begin{figure}
    \centering
    \resizebox{\linewidth}{!}{
   \begin{quantikz}
& \gate{H} & \ctrl{1}  &\qw & \qw & \qw & \qw & \qw & \ctrl{1} & \gate{H} & \qw  \\
& \gate{R_x(\frac{-\pi}{2})} & \targ{} &\ctrl{1} & \qw & \qw & \qw & \ctrl{1} & \targ{} & \gate{R_x(\frac{\pi}{2})} & \qw \\
\lstick{$\vdots$}& \gate{...} & \qw  &\targ{} & \ctrl{1} &\qw & \ctrl{1} & \targ{} & \qw & \gate{...} & \qw \\   
& \gate{\mathbb{1}} & \qw &\qw & \targ{} & \gate{R_{z}(\theta)} & \targ{} & \qw & \qw & \gate{\mathbb{1}} & \qw 
\end{quantikz}}
    \caption{Pauli Gadget circuit for implementing $e^{-i\frac{\bm\theta}{2} (XY...Z)}$}
    \label{fig:pauligadget}
\end{figure}

The trotterized and truncated UCC operator can be translated into quantum gates in two steps. The first step is to map the $\hat{a}^\dagger$ and $\hat{a}$ of excitation operators into Pauli strings (tensor products of Pauli matrices, acting on several qubits) as described in section \ref{subsection:mapping}. Then, each exponentiated Pauli string can be translated into a Pauli gadget \cite{nielsen2010}, as shown in figure \ref{fig:pauligadget}. The ansatz parameters $\bm\theta$ are the rotation angles in the gates $R_z$, denoting single qubit rotations around the $Z$ axis.
The sequence of Pauli gadgets gathered into one single circuit constitutes the UCC ansatz.
\smallskip

\subsubsection{Ordering of excitation operators}\label{uccorder}

The Trotterized form of UCC is a product of non-commuting terms, making the ordering of excitation operators an important hyper-parameter of the ansatz. In Ref~\citenum{Evangelista2019}, authors have proposed a universal ordering of the excitations allowing us to reach any fermionic state . Their method is employed in this study. Considering a single determinant reference state $\ket{\Phi_0}$ (chosen as the Hartree--Fock wavefunction here), we iterate through the occupied indices $j$ of $\ket{\Phi_0}$. For each index, all single excitations involving index $j$ ($e^{\theta_j^a \hat{a}^\dagger_a \hat{a}_j}$) are added to the ansatz, followed by all double excitations with index $j$ ($e^{\theta_{ij}^{ab} \; \hat{a}^\dagger_a\hat{a}^\dagger_b \hat{a}_i\hat{a}_j}$) and so on for all excitations of higher rank involving orbital $j$. This procedure is repeated for all occupied indices $j$ of the reference state, eventually adding all excitations of the UCC ansatz to the circuit.
\bigskip

\subsection{Fermion-to-qubit mapping}
\label{subsection:mapping}

Second quantized fermionic operators can be mapped to qubit operators, implementable on a quantum circuit.

Different mapping schemes are available, the most common being the Jordan--Wigner (JW) \cite{JWmap} and Bravyi--Kitaev (BK) \cite{BKmap} transformations. In the second quantized formalism, fermionic operators are expressed as sums of creation and annihilation operators. Fermion-to-qubit mapping is a systematic formula to translate creation and annihilation operators into Pauli strings.\newline
Any second quantized operator :
\begin{equation}\label{eq:2ndqop}
    \hat{O} = \sum_{pq} o_{pq}\hat{a}_p^\dagger\hat{a}_q + \sum_{pqrs} o_{pqrs}\hat{a}_p^\dagger\hat{a}_q^\dagger\hat{a}_r \hat{a}_s + ...
\end{equation}
can be mapped to :
\begin{equation}\label{eq:pstrings}
    \hat{O} = \sum_i o_i \hat{P}_i
\end{equation}
with $\hat{P}_i$ being Pauli strings and $o_i$ scalars.

In this work, Jordan--Wigner mapping was employed.

\subsubsection{Jordan--Wigner Mapping}
In this formalism, each qubit represents a fermionic state (a spin-orbital for molecules), with the qubit state $\ket{0}$ corresponding to an unoccupied state, and $\ket{1}$ to an occupied state. The creation and annihilation operators are mapped using the transformation in equation (\ref{eq:JW}) for a $N$-qubit register corresponding to $N$ electronic spin-orbitals.

\begin{equation}\label{eq:JW}
    \begin{aligned}
    \hat{a}_j^\dagger \leftrightarrow Z_1 \otimes Z_2 \otimes ... \otimes Z_{j-1} \otimes \sigma^+_j \otimes \mathbb{1}_{j+1} ... \otimes \mathbb{1}_{N} \\
    \hat{a}_j \leftrightarrow Z_1 \otimes Z_2 \otimes ... \otimes Z_{j-1} \otimes \sigma^-_j \otimes \mathbb{1}_{j+1} ... \otimes \mathbb{1}_{N}
    \end{aligned}
\end{equation}

with gates $$Z_j=\begin{bmatrix}
1 & 0\\
0 & -1
\end{bmatrix}$$ 
$$\sigma^+_j=\begin{bmatrix}
0 & 0\\
1 & 0
\end{bmatrix} = \frac{X-iY}{2}$$
$$\sigma^-_j=\begin{bmatrix}
0 & 1\\
0 & 0
\end{bmatrix} = \frac{X+iY}{2}$$ applied to qubit $j$. 

The $\sigma^+$ and $\sigma^-$ gates act as qubit creation and annihilation operators, while the $Z$ gates are required to conserve the anti-commutation relations : 
\begin{equation}\label{eq:fermioncommut}
   \{a^{\,}_i, a^\dagger_j\} \equiv a^{\,}_i a^\dagger_j +a^\dagger_j a^{\,}_i = \delta_{i j}.
\end{equation}

\subsection{VQE procedure}

\begin{figure}[h]
    \centering
    \includegraphics[width=\linewidth]{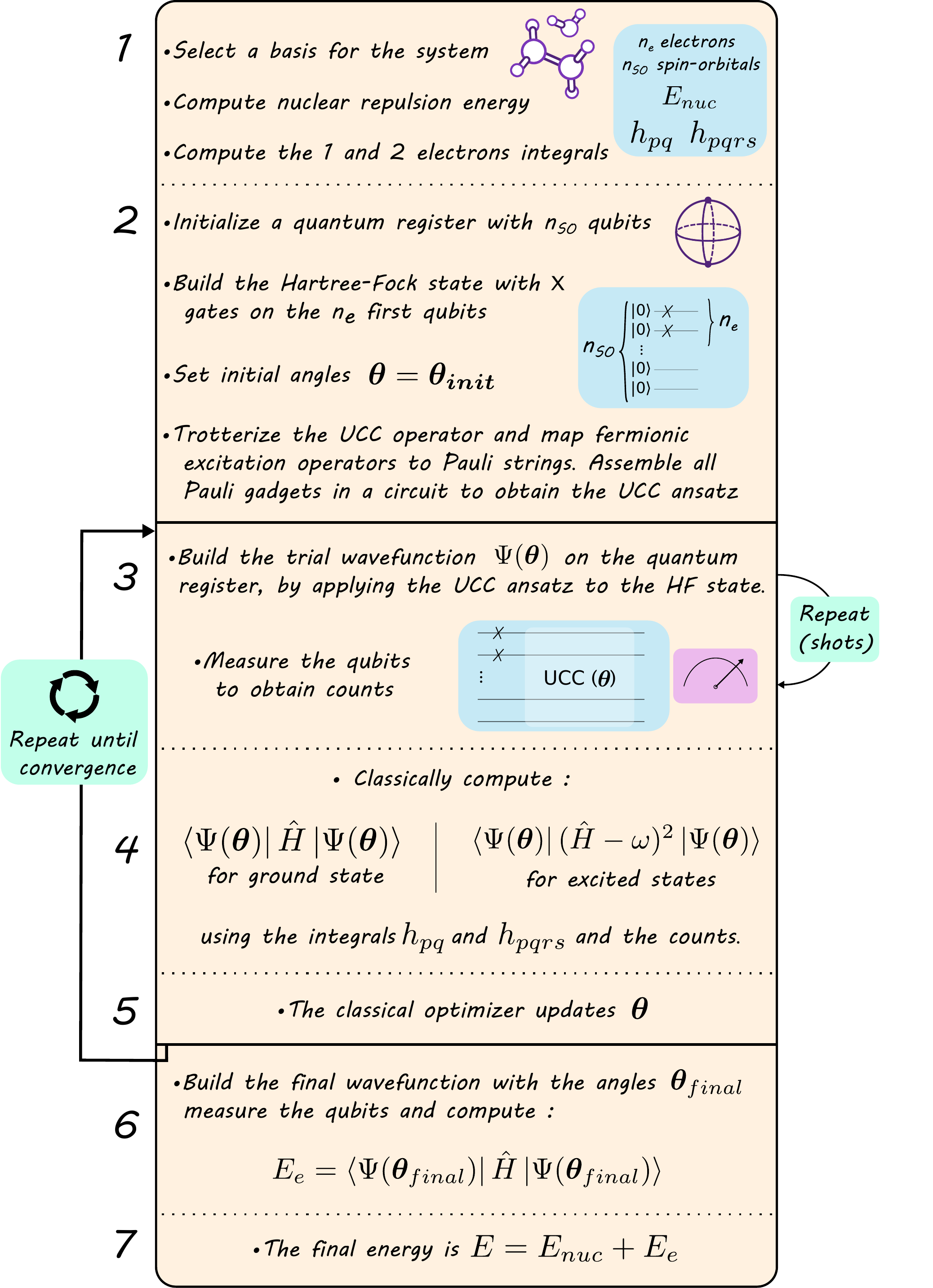}
    \caption{Summary of the pre-processing steps and VQE algorithm to obtain a molecule's energy.}
    \label{fig:vqeproc}
\end{figure}

The procedure implemented in this work is summarized in figure \ref{fig:vqeproc}.

\section{Excited States}

The standard VQE algorithm applied to molecular systems allows one to compute the ground-state of the electronic wavefunction. It is not primarily designed to compute excited states, as it relies on the minimisation of the average energy.
Several approaches have been proposed to reach excited states with quantum algorithms, including Quantum Subspace Expansion (QSE) \cite{McClean2017}, Variational Quantum Deflation (VQD) \cite{Higgott2019} similar to Orthogonally Constrained VQE (OC-VQE) \cite{Lee2018,Jones2019}, witnessing eigenstates (WAVES) \cite{Santagati2018} or Quantum Equation of Motion \cite{Ollitrault2020,Asthana2023}. 

The Folded Spectrum (FS) method has also been reported in the literature \cite{Cao2019}, but the presence of a quadratic term in $\hat{H}$ is regarded prohibitive and it is expected to scale as $\mathcal{O}(N^8)$ relative to system size \cite{Lee2018}. To the best of our knowledge, no extensive study of this method has been reported.

\subsection{Folded Spectrum method}

The principle of the FS method is to minimise the expectation value of the FS operator $(\hat{H}-\omega)^2$ instead of the Hamiltonian $\hat{H}$, with $\omega$ an arbitrary target energy.

This method is also known as state-specific variance minimisation in the Quantum Monte Carlo (QMC) literature, where it has been actively employed and studied for many years \cite{Umrigar1988,Hanscam2022,Otis2023}.

Let $\ket{\Psi}$ be an eigenstate of the Hamiltonian $\hat{H}$. It satisfies the time-independent Schrödinger equation :
\begin{equation}\label{eq:Schrodinger}
    \hat{H} \ket{\Psi} = E \ket{\Psi}.
\end{equation}

The linearity of the Schrödinger equation allows one to write equation (\ref{eq:FS}) for all $\ket{\Psi}$ eigenstates of $\hat{H}$, $E$ the associated eigenvalues, and $\omega$ an arbitrary scalar.

$$\forall \; \ket{\Psi} \text{ such that } \hat{H} \ket{\Psi} = E \ket{\Psi},$$
   $$ \forall \; \omega :$$
\begin{equation}\label{eq:FS}
    (\hat{H}-\omega)^2 \ket{\Psi} = (E-\omega)^2 \ket{\Psi}
\end{equation}

\begin{figure}[ht]
    \centering
    \includegraphics[width=\linewidth]{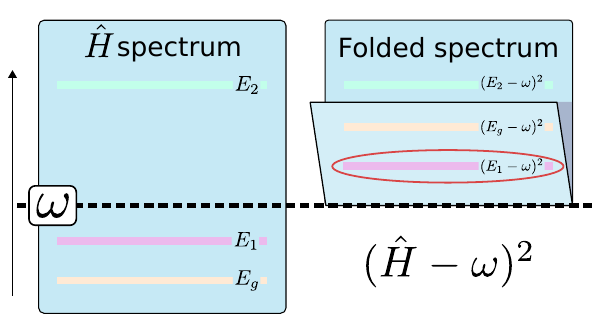}
    \caption{Illustration of Folded Spectrum method. Colored lines represent eigenstates along a vertically ascending axis. The eigenspectrum of $\hat{H}$ (left) is \emph{folded} around $\omega$ in the spectrum of $(\hat{H}-\omega)^2$ (right), causing its eigenvalues to reorder. The lowest eigenvalue of the folded spectrum, circled in red, is the excited eigenstate of $\hat{H}$ originally closest to $\omega$.}
    \label{fig:FS}
\end{figure}

The FS operator $(\hat{H}-\omega)^2$ and the Hamiltonian $\hat{H}$ share the same eigenstates but with a reordering in the eigenvalues (corresponding to a \emph{fold} around $\omega$) \cite{FS1994}. The lowest lying eigenstate of the folded operator is the one with an energy $E_i$ closest to $\omega$ (see figure \ref{fig:FS}).

By minimising the expectation value of the FS operator $(\hat{H}-\omega)^2$, one can find an eigenstate of $\hat{H}$ such that $(E_i-\omega)^2$ is minimal, and thus obtain an excited state of the Hamiltonian, close to the target energy $\omega$. In practice, we perform expectation value minimization with a VQE procedure, and the obtained wavefunction is an approximation of the true eigenstate given by the ansatz.
The cost function of interest is :

\begin{equation}
    F(\bm\theta) = \bra{\Psi(\bm\theta)} (\hat{H}-\omega)^2 \ket{\Psi(\bm\theta)}.
\end{equation}

All excited states of the Hamiltonian may be obtained by modifying the parameter $\omega$ over a wide enough range of energies.\\

One major limitation of the FS method is that it requires a squared Hamiltonian, containing a large number of fermionic terms and thus of Pauli strings compared to the Hamiltonian itself. However, by using a Pauli reduction and grouping procedure as described in section \ref{subsec:pauligroup}, the number of required measurements can be considerably reduced.

\section{Methods}

\subsection{Computing expectation values}\label{subsec:expvalcomp}

The expectation value of quantum operators represented by Pauli strings (as in equation (\ref{eq:pstrings})) can be decomposed as shown in equation (\ref{eq:expvaldecompo}), with $\hat{P}_i$ Pauli strings and $o_i$ scalar coefficients.

\begin{equation} \label{eq:expvaldecompo}
 \begin{aligned}
    \langle \hat{O} \rangle = \bra{\Psi}\hat{O}\ket{\Psi} = \bra{\Psi}\sum_i o_i\hat{P}_i\ket{\Psi} \\ = \sum_i  o_i \bra{\Psi}\hat{P}_i\ket{\Psi} = \sum_i  o_i \langle \hat{P}_i \rangle 
\end{aligned}
\end{equation}

The expectation value of $\hat{O}$ can be computed by classically summing the expectation values of each Pauli string $\langle\hat{P}_i\rangle$ weighted by the pre-computed coefficients $o_i$. In molecular Hamiltonians the coefficients $o_i$ are typically denoted by $h_i$ and are formed through linear combinations of one-body and two-body integrals.\bigskip

Once the trial state $\ket{\Psi}$ is prepared on the quantum register, we measure the qubits and repeat the state preparation and measurement procedure several times (\emph{shots}). Ultimately, we obtain some \emph{counts} that form estimates for $\ket{\Psi}$ populations.
Qubit measurements are usually performed in the computational basis denoted $\{\Phi_i\}_{i=1}^{2^n}$, corresponding to the values $0$ or $1$ for each of the $n$ qubits : $\{\Phi_i\} = \{\ket{0...00},\ket{0...01},...,\ket{1...11}\}$ in binary order. 
\smallskip
In this basis, the spectral decomposition of $\ket{\Psi}$ is :

\begin{equation}
    \ket{\Psi} = \sum_{i=1}^{2^n}\alpha_i\ket{\Phi_i},
\end{equation}
with
\begin{equation}
    \alpha_i = \braket{\Phi_i}{\Psi}.
\end{equation}

Measurable quantities are the populations for each basis vector of the computational basis :

\begin{equation}
    \{|\alpha_i|^2\}_i^{2^n}.
\end{equation}

Note that the counts provide \emph{estimates} of the populations, due to finite sampling. The final precision $\epsilon$ in the results is directly correlated with the number of shots taken $s$ as $\epsilon \sim \frac{1}{\sqrt{s}}$ (see Appendix \ref{app:precision}). Results can be made arbitrarily close to the theoretical value by increasing the number of shots, but this can lead to considerable computing time. To reduce the number of quantum measurements, a Pauli grouping routine can be used, as discussed in the next section.

The derivation of expectation values from the counts results is explained below for diagonal and non-diagonal operators.

\subsubsection{Diagonal Operators}
The expectation value of diagonal Pauli operators in the computational basis (i.e. tensor products of $I$ and $Z$) can be directly computed from the counts.
Such operators can be decomposed in the computational basis as a sum of projectors :
\begin{equation}
    \hat{P}_\mathrm{diag} = \sum_i \lambda_i \ket{\Phi_i}\bra{\Phi_i}
\end{equation}

\noindent with $\lambda_i$ corresponding to the eigenvalues of the operator, namely $+1$ or $-1$ for products of $I$ and $Z$ Pauli operators. \smallskip

The expectation value of $\hat{P}_\mathrm{diag}$ is therefore :

\begin{equation}\label{eq:expval}
 \begin{aligned}
     \bra{\Psi}\hat{P}_\mathrm{diag}\ket{\Psi} = \bra{\Psi}\left(\sum_i \lambda_i\ket{\Phi_i}\bra{\Phi_i} \right)\ket{\Psi} = \\ \sum_i  \lambda_i \bra{\Psi}\ket{\Phi_i}\bra{\Phi_i}\ket{\Psi}= \sum_i  \lambda_i \left|\bra{\Phi_i}\ket{\Psi}\right|^2.
 \end{aligned}
\end{equation}

Consequently :
\begin{equation}
    \langle \hat{P}_\mathrm{diag} \rangle = \sum_i \lambda_i \; |\alpha_i|^2
\end{equation}
is directly accessible from the quantum measurement, by classically summing the count results weighted by the eigenvalues of the Pauli operator.

Note that all diagonal Pauli operators can be evaluated from the same counts measurement, as the $\lambda_i$ coefficients are treated classically.

\subsubsection{Non-diagonal operators}\label{par:expvalnd}
Given a non-diagonal operator $\hat{P}$ in the qubit basis, it is always possible to find an appropriate basis change to diagonalize it.
$X$ and $Y$ Pauli matrices are non-diagonal in the computational basis, but they can be diagonalized using the following basis changes:
\begin{equation}X = H^\dagger ZH\end{equation}
\begin{equation}Y = R_x\left(\frac{-\pi}{2}\right) \; Z \; R_x\left(\frac{\pi}{2}\right).\end{equation}

Therefore, for a Pauli string $\hat{P}$, it is possible to find a rotated Pauli string $\tilde{P}$ diagonal in the qubit basis, using these basis changes. In general, one can write :
\begin{equation} \label{eq:rot}
\hat{P} = R^\dagger \tilde{P} R
\end{equation}
with
\begin{equation}
    \tilde{P} = \sum_i \tilde{\lambda}_i \ket{\Phi_i}\bra{\Phi_i}.
\end{equation}

The expectation value of $\hat{P}$ is then :

\begin{equation} \langle \hat{P} \rangle = \expval{\hat{P}}{\Psi} = \expval{R^\dagger \tilde{P} R}{\Psi} = \expval{\tilde{P}}{\tilde{\Psi}},\end{equation}

with \begin{equation}\ket{\tilde{\Psi}} = R \ket{\Psi}.\end{equation}

Finally,
\begin{equation}\label{eq:rotexpval}
   \langle \hat{P} \rangle = \sum_i  \tilde{\lambda}_i \left|\bra{\Phi_i}\ket{\tilde{\Psi}}\right|^2.
\end{equation}

To evaluate the expectation value of a non-diagonal Pauli string $\hat{P}$, it is therefore necessary to apply a post-rotation gate $R$ to the quantum state $\ket{\Psi}$, which creates a state $\ket{\tilde{\Psi}}$ in a basis where the Pauli string is diagonal.
In practice, the additional post-rotation operator is built with single-qubit gates added after the ansatz. A Hadamard is applied to the qubits where the Pauli operator is $X$, and a $R_x\left(\frac{\pi}{2}\right)$ is applied to the qubits where it is $Y$.

The expectation value of a non-diagonal $\hat{P}$ is then computed similarly to a diagonal Pauli string, using the rotated state $\ket{\tilde{\Psi}}$ (equation (\ref{eq:rotexpval})).

\subsection{Pauli strings reduction and grouping}\label{subsec:pauligroup}

 The number of Pauli strings in the Hamiltonian scales polynomially with the system size, and naively the FS operator can contain up to the square of this number. Evaluating each term one by one can lead to a very large number of measurements, which lowers the potential advantage of the quantum algorithm.\smallskip

\subsubsection{Pauli reduction}

When computing the FS operator $(\hat{H}-\omega)^2$, the number of Pauli strings primarily obtained is approximately the square of the number of terms in $\hat{H}$. It is possible to simplify and reduce this sum by using the commutation and anti-commutation relations between Pauli matrices \cite{Aulicino2021,Claudino2021}.
In Ref~\citenum{Suchsland2021}, authors studied a collection of systems of increasing sizes and concluded that the actual number of Pauli strings in $\hat{H}^2$ after Pauli reduction has an effective scaling below $\mathcal{O}(N^6)$ instead of the expected $\mathcal{O}(N^8)$ with $N$ the number of spin-orbitals. This result can be extended to our work : the number of terms in the FS operator has a much more favorable scaling with respect to the system size due to Pauli reduction. More formal analyses are required to consolidate this result and assess the feasibility of the FS method for larger systems.

\subsubsection{Pauli grouping}

\begin{figure}
    \centering
    \includegraphics[width=\linewidth]{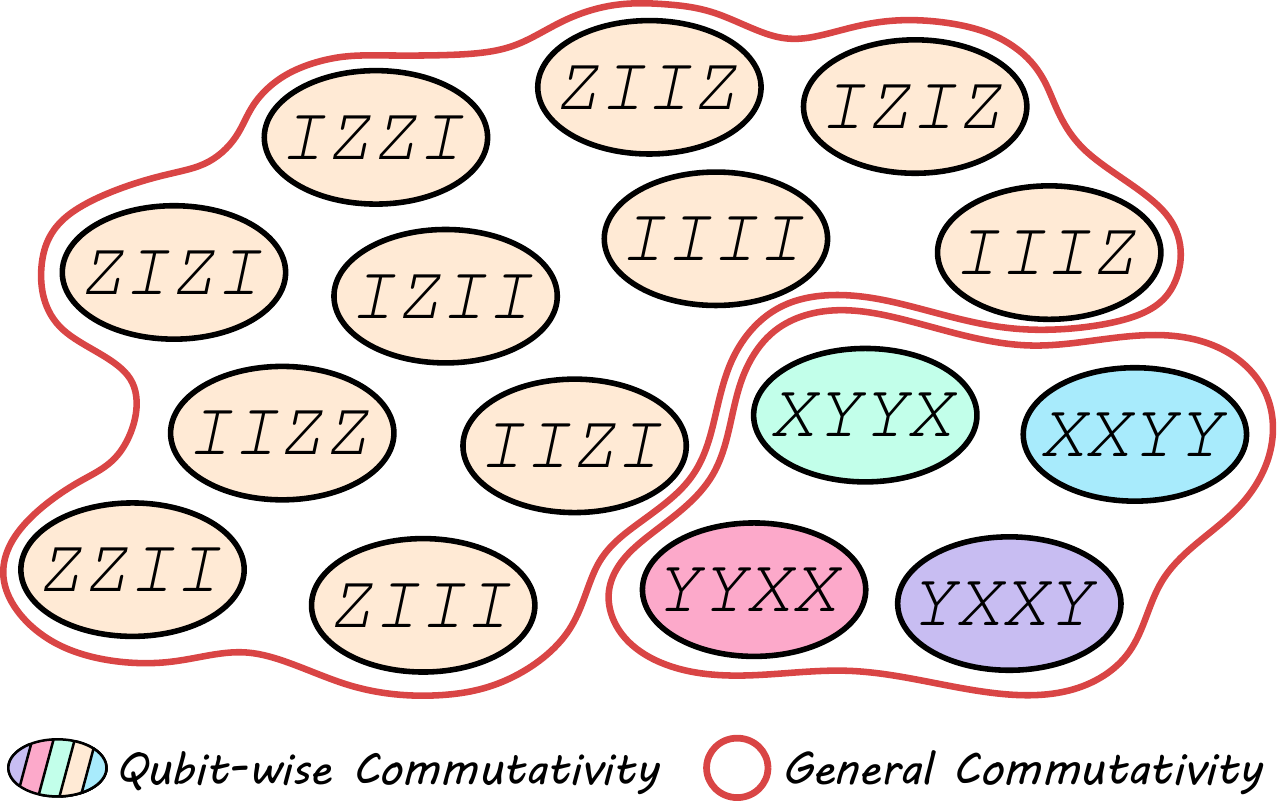}
    \caption{Example of Pauli grouping for the 15 Pauli strings in $\ce{H2}$ Hamiltonian. Colors show the QWC partitioning that reduces the number of evaluations to 5, and red circles show the GC partitioning that only require 2 evaluations.}
    \label{fig:Pgroup}
\end{figure} 

 To further reduce the number of quantum evaluation required, one can partition the operators into groups of simultaneously diagonalizable Pauli strings \cite{Kandala2017IBM,McClean2016}.
All the Pauli strings in the same group can have their expectation values determined with a single quantum evaluation, by adding a classical post-processing step. 
In the formalism of section \ref{par:expvalnd}, it means that all Pauli strings in the group share the same post-rotation $R$ in equation (\ref{eq:rot}). \\ Formally speaking, a group of operators is simultaneously diagonalizable if and only if the operators commute \cite{Horn2012-jp}. This reduces the problem to identifying groups of commuting Pauli strings in the qubit Hamiltonian or FS operator. In particular, we want to find a partitioning with a minimal number of groups, leading to a minimal number of quantum evaluations.

Two distinct definitions of commutation can be considered to partition the Pauli strings: qubit-wise commutativity (QWC) or general commutativity (GC) \cite{Gokhale2020}. The former defines that two Pauli strings commute if the Pauli matrices commute at each index. For instance the group \{$IX$, $XX$, $XI$\} is QWC since all Pauli matrices for qubit 1 commute, and equally for qubit 2.  \\
General commutativity is fulfilled if the two Pauli strings commute, regardless of the single-qubit case. The group \{$XX$, $YY$, $ZZ$\} is GC although none of the pairs is QWC. The general rule is that each pair must fail to commute at an even number of indices.
QWC is, in fact, a special case of GC where the strings fail to commute at 0 indices. Figure \ref{fig:Pgroup} shows an example of QWC and GC partitioning for the electronic Hamiltonian of $\ce{H2}$. \\
Finding the optimal Pauli partitioning (in QWC or GC) is equivalent to a graph partitioning problem known as the minimum clique cover problem \cite{Golumbic2004-fq}, and it is NP hard \cite{Miller1972-vj}. Efficient heuristic algorithms to find a good Pauli partitioning are therefore essential to tend toward scalability for the FS method \cite{Verteletskyi2020,Huggins2021}.

In this work, we use QWC Pauli partitioning as it is more straightforward to implement. We present some results using Jordan--Wigner qubit mapping in figure \ref{fig:neval}, and table \ref{tab:pgroup} in appendix \ref{app:pgroup} reports some examples for both J--W and B--K transformations. The number of quantum evaluations is systematically decreased by grouping the Pauli strings. As expected, more evaluations are required for the FS operator compared to the Hamiltonian for the same system. Additional results on Hamiltonian grouping for other transformations and systems can be found in Refs ~\citenum{Izmaylov2019,Verteletskyi2020,Yen2023}.
As shown in Ref~\citenum{Gokhale2020}, GC partitioning is more efficient than QWC, and it would lead to even fewer quantum measurements for the FS method.

\begin{figure}
    \centering
    \includegraphics[width = \linewidth]{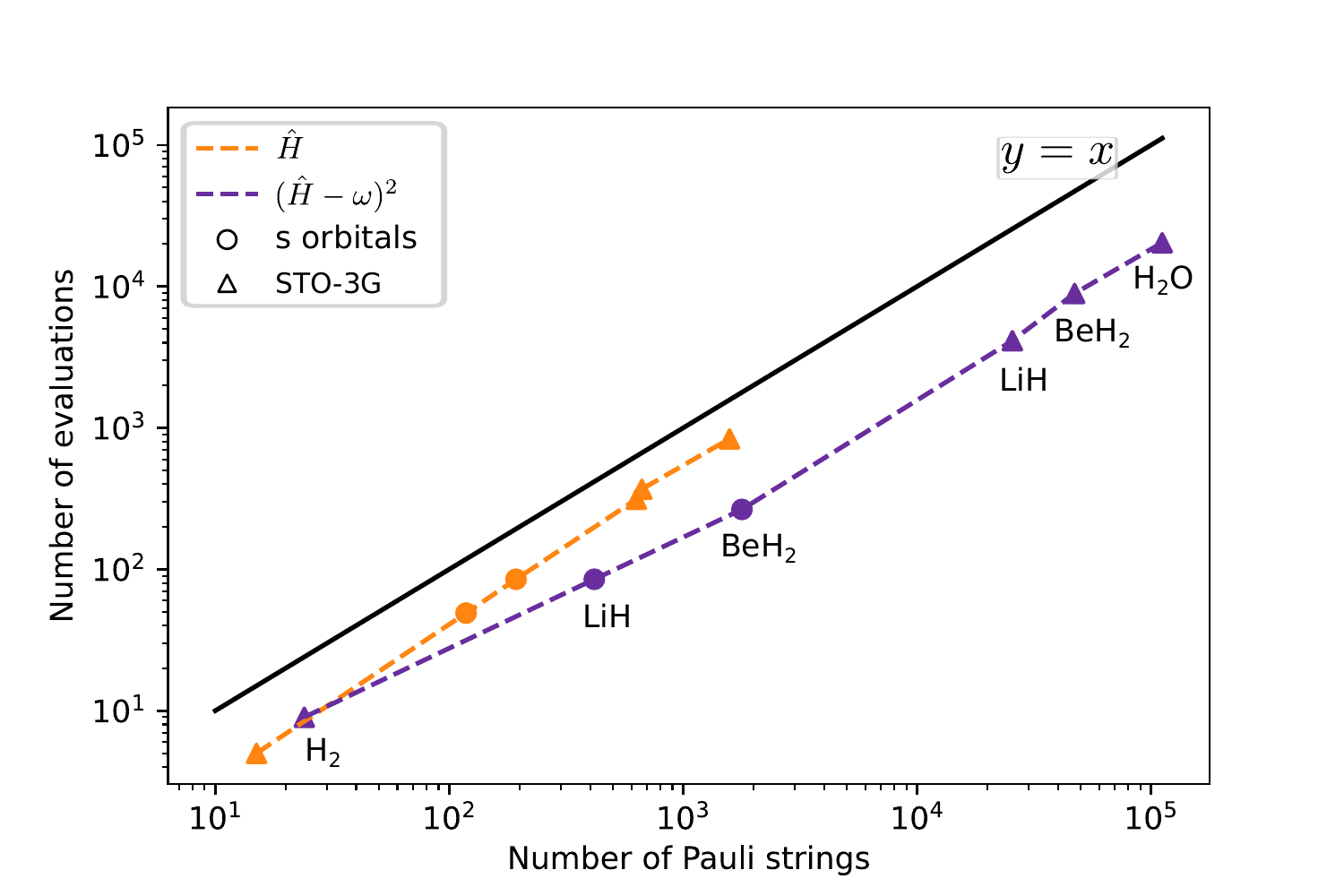}
    \caption{Number of evaluations needed (number of QWC groups) compared to the number of Pauli strings for the Hamiltonian and the FS operator, under JW transformation. Several systems are displayed (labels are in matching order) in STO-3G basis (triangles) or in a minimal basis with only s orbitals (circles). The black line shows the number of evaluations needed if measuring the Pauli strings one by one.}
    \label{fig:neval}
\end{figure} 

\subsubsection{Measurement cost}
Pauli grouping for molecular Hamiltonians shows interesting features that may be extended to the FS operator.
Electronic Hamiltonians under certain fermion-to-qubit transformations (such as JW mapping described in section \ref{subsection:mapping}) have the convenient property of including a large number of diagonal Pauli strings (with only $I$ and $Z$ operators). In fact, the product of $\sigma^+$ and $\sigma^-$ operators defined in equation (\ref{eq:JW}) can be decomposed as shown in equation (\ref{eq:Pgrouping}).

\begin{equation}\label{eq:Pgrouping}
\begin{aligned}
    \sigma^+\times\sigma^-= \frac{I-Z}{2} \\
    \sigma^-\times\sigma^+= \frac{I+Z}{2}.
\end{aligned}
\end{equation}

Therefore, all terms in the Hamiltonian involving a creation and an annihilation operator for the same spin-orbital will be mapped to a diagonal Pauli string in the qubit basis. These terms are the one-body terms with $p=q$ ($h_{pp}$) and the two body terms with $p=r$ and $q=s$ $h_{prpr}$ or $p=s$ and $q=r$ ($h_{prrp}$). For a Hamiltonian describing $n$ electrons in $N$ spin-orbitals, there are $N$ diagonal one-body operators and $\binom{N}{2}$ diagonal two-body operators. All these diagonal Pauli strings can therefore be grouped together and evaluated simultaneously. \smallskip
 
Additionally, asymptotically dominant terms in the molecular Hamiltonian are two-electrons operators of the form $h_{pqrs}\hat{a}^\dagger_p\hat{a}^\dagger_q\hat{a}_r\hat{a}_s$ with $p\neq q\neq r \neq s$ \cite{Gokhale2020}. The number of such terms in a molecular Hamiltonian describing $N$ spin-orbitals scales as $\binom{N}{4} \sim \mathcal{O}(N^4)$. \newline
Under Jordan--Wigner mapping, two of these terms with disjoint indices, namely 
$\hat{a}^\dagger_p\hat{a}^\dagger_q\hat{a}_r\hat{a}_s$ and $\hat{a}^\dagger_i\hat{a}^\dagger_j\hat{a}_k\hat{a}_l$ with $\{p,q,r,s\} \cap \{i,j,k,l\} = \emptyset$
\noindent involve disjoint qubits for $X$ and $Y$ gates, and authors of Ref~\citenum{Gokhale2020} demonstrated that they commute. They showed that using the Baranyai's graph coloring theorem \cite{Baranyai1974}, it is possible to partition these $\binom{N}{4} \sim \mathcal{O}(N^4)$ terms into $\binom{N-1}{3} \sim \mathcal{O}(N^3)$ groups such that the operators within each set have disjoint indices, and therefore commute.\newline
In other words, instead of measuring each of the $\mathcal{O}(N^4)$ terms individually, one can perform $\mathcal{O}(N^3)$ measurements only to compute the expectation value of the asymptotically dominant Pauli strings in the molecular Hamiltonian.\smallskip

Similarly to the Hamiltonian, the FS operator can be partitioned into commuting groups to reduce the number of measurements. Here, we present empirical data on the effect of Pauli grouping for the FS operator. Future studies could aim to establish analytical results on the scaling of the number of evaluations needed for the FS method.

\subsection{Reference state}
When computing the molecular ground-state, the Hartree--Fock (HF) determinant can often be used as a reference because of the significant overlap between the HF state and the FCI electronic ground-state. 
This is not generally true for excited states. In this case an excited single determinant or a superposition of two or more determinants having overlap with the target wavefunction can be used as a reference.

In this study, we have selected relevant references for all electronic excited states by exciting the ground HF determinant with single and double excitations and symmetrizing the spin function when necessary. This procedure can be generalised to larger systems, but we expect that more sophisticated reference states may be required for molecules with strong multi-reference character. This question remains an essential challenge for the scalability of the FS method.
The reference states used in this paper are specified in the results section.

\section{Noise robustness and error mitigation}

The presented algorithm is based on the Variational Quantum Eigensolver, which is designed to be amenable on near-term quantum devices. To assess the feasibility of our method on noisy devices, we evaluated its noise robustness by including noise models in our simulations and using zero noise extrapolation (ZNE) and state preparation and measurement (SPAM) error mitigation techniques \cite{MitigCai}.

\subsection{Noise Model}

We designed a tunable and realistic noise model based on the models provided for IBM quantum devices \cite{Qiskit}. This model includes: \begin{itemize}
    \item \textbf{gate errors} consisting of a depolarizing channel characterized by one and two qubit gates error rates $p_1$ and $p_2$ respectively, followed by thermal relaxation and dephasing processes, driven by $T_1$ and $T_2$ characteristic times applied for the gate lengths $t_\mathrm{gate1}$ and $t_\mathrm{gate2}$ for one and two qubits gate.
    \item \textbf{readout error} characterized by an error probability $p_{\mathrm{SPAM}}$ for each qubit.
\end{itemize}

This model can be considered realistic for quantum computers, provided that the parameters are adjusted to the calibration data of the real device. However, it excludes noise sources such as state leakage or cross-talk which are more complex to model. This approximation is widely adopted for its simplicity, but may not be suitable for some devices where leakage or cross-talk effects are not negligible \cite{Georgopoulos2021}.

Parameters $T_1$, $T_2$, $t_\mathrm{gate1}$, $t_\mathrm{gate2}$ $p_1$, $p_2$ and $p_{\mathrm{SPAM}}$ are provided for IBM machines. Some typical values for the present devices are given in Table \ref{tab:noiseparams} \cite{ibmQuantum}. 

\begin{table}[h]
    \centering
    \begin{tabular}{c|ccccccc}
    \hline 
         & $T_1$  & $T_2$  & $t_\mathrm{gate1}$ & $t_\mathrm{gate2}$ & $p_1$ & $p_2$ & 
         $p_{\mathrm{SPAM}}$ \\ 
         & ($\mu$s) & ($\mu$s) & (ns) & (ns) & & &  \\ \hline 
        $\lambda = 1$ & 290 & 145 & 35 & 300 & $10^{-4}$ & $10^{-3}$ & $10^{-2}$  \\
 \hline
        $\lambda = 0.4$ & 924 & 461 & 14 & 120 & 4$\times10^{-5}$ & 4$\times10^{-4}$ & 4$\times10^{-3}$ \\ \hline 
        $\lambda = 0$ & $\infty$ & $\infty$ & 0 & 0 & 0 & 0 & 0 \\ \hline

    \end{tabular}
    \caption{Values of the noise parameters for different values of the scale factor $\lambda$. $\lambda = 1$ corresponds to the order of magnitude of the experimental values in IBM machines \cite{ibmQuantum}, and $\lambda = 0$ is the ideal value in a noiseless device. In practice, ideal values were fixed at $T_{1\infty} = 2000 \mu s$ and $T_{2\infty} = 1000 \mu s$.}
    \label{tab:noiseparams}
\end{table}

To vary the noise level, we scaled the parameters according to a scale factor $\lambda$ between their experimental value ($\lambda=1$) and their ideal value ($\lambda$=0). The scaling is performed exponentially for $T_1$ and $T_2$ and linearly for all other parameters. Ideal values for $T_1$ and $T_2$ were fixed at two orders of magnitude above the total length of the circuit, that is, $T_{1\infty} = 2000 \mu s$ and $T_{2\infty} = 1000 \mu s$ here. 

\subsection{State Preparation and Measurement mitigation}\label{subsec:mitig}

The SPAM technique aims to mitigate errors introduced during the state preparation and measurement stages \cite{REM,Nation2021SPAM}. A confusion matrix $\mathcal{A}$ is measured by running small calibration circuits on the quantum processor. $\mathcal{A}$ represents the noise channel corresponding to the probability that the state preparation or measurement outcome will be incorrect for each qubit. This matrix is then inverted, and the inverse channel $\mathcal{A}^{-1}$ is classically applied to the next experiments as a post-processing step, thus obtaining quasi-probabilities with mitigated SPAM error. The nearest probability distribution is then selected as the mitigated measurement result. This method assumes that the SPAM noise remains constant over multiple experiments close in time on the same device. The calibration circuits should be run regularly to make this assumption reasonable. A more detailed description of the SPAM error channel can be found in Ref~\citenum{Georgopoulos2021}.
In our implementation, the confusion matrices are directly extracted from the noise model's parameters $p_{\mathrm{SPAM}}$ for each qubit.

\subsection{Zero Noise Extrapolation}

Zero noise extrapolation (ZNE) is a technique that aims to mitigate the effect of noise when evaluating expectation values on QPUs \cite{GiurgicaTiron2020}. In ZNE, the hardware noise level is represented by a parameter $\gamma$, with $\gamma=1$ corresponding to the actual noise level of the quantum computer, $\gamma >1$ being a noisier hardware and vice versa.

The principle of ZNE is to intentionally increase the noise level ($\gamma = 3,5,7...$), and to evaluate the same expectation value $E_\gamma$ for the different values of $\gamma$. The points obtained are plotted on a $E_\gamma$ vs $\gamma$ curve and fitted with an analytic model. The model provides an extrapolated value at $\gamma=0$ that is retained as the mitigated result $E_0$, which should be an approximate of the ideal value $E$. 

In our implementation, ZNE was employed to mitigate the summed expectation values of each Pauli groups separately. For $\ce{H2}$, the Folded Spectrum operator contains 24 Pauli strings that can be partitioned using general commutativity into two commuting groups $\mathcal{G}_1$ and $\mathcal{G}_2$ :
\begin{equation}
\begin{aligned}
     \mathcal{G}_1 = & \{IIII,IIIZ,IIZI,ZZZZ,IIZZ, IZIZ, \\
                     &  ZIII,ZIIZ,ZIZI,IZII, IZZI,IZZZ, \\ 
                     & ZIZZ,ZZZI,ZZII,ZZIZ \} \\
    \mathcal{G}_2 = & \{ XYXY,YYYY,XXYY,YXXY,\\ 
    & XYYX,YYXX,XXXX,YXYX\}.
\end{aligned}
\end{equation}

For this system, each evaluation of $\langle(\hat{H}-\omega)^2\rangle$ requires measuring two different circuits, one for each group. A ZNE procedure can be used to mitigate the noise on the summed expectation value of each commuting group. 

To increase the noise level $\gamma$, we used a method named \textbf{unitary folding} that consists of replacing a unitary gate sequence $\mathcal{G}$ by a folded version (for example, $\mathcal{G}\mathcal{G}^\dagger\mathcal{G}$). The logical operations of $\mathcal{G}$ and its folded versions are the same as $\mathcal{G}\mathcal{G}^\dagger = \mathds{1}$, but the effective number of gates in $\mathcal{G}\mathcal{G}^\dagger\mathcal{G}$ is three times greater, corresponding to $\gamma = 3$. Further folding is performed to implement $\gamma = 5,7,9...$. On noisy devices, folding the circuits effectively corresponds to scaling the gate noise. Thus, this method allows us to artificially implement different values of $\gamma$ by increasing the circuit depth. This method has the advantage of being very general since it can work for any unitary circuit.

A quadratic model in equation (\ref{eq:quad}) was selected to fit and extrapolate the $E_{\gamma}(\gamma)$ curve.
\begin{equation}\label{eq:quad}
E_{\gamma} = a \gamma^2 + b \gamma + c
\end{equation}

ZNE relies on the assumption that folding the circuits corresponds to increasing the noise level by $\gamma$ overall, which implies that gate noise is the main source of noise. This assumption is not always valid, and in particular, SPAM error is not scaled with circuit folding, which can make the extrapolation process erroneous. To alleviate this limitation, we employ ZNE mitigation in conjunction with SPAM mitigation, such that in theory SPAM errors are negligible in the ZNE fitted data.

\begin{figure}[h]
    \centering
    \includegraphics[width=\linewidth]{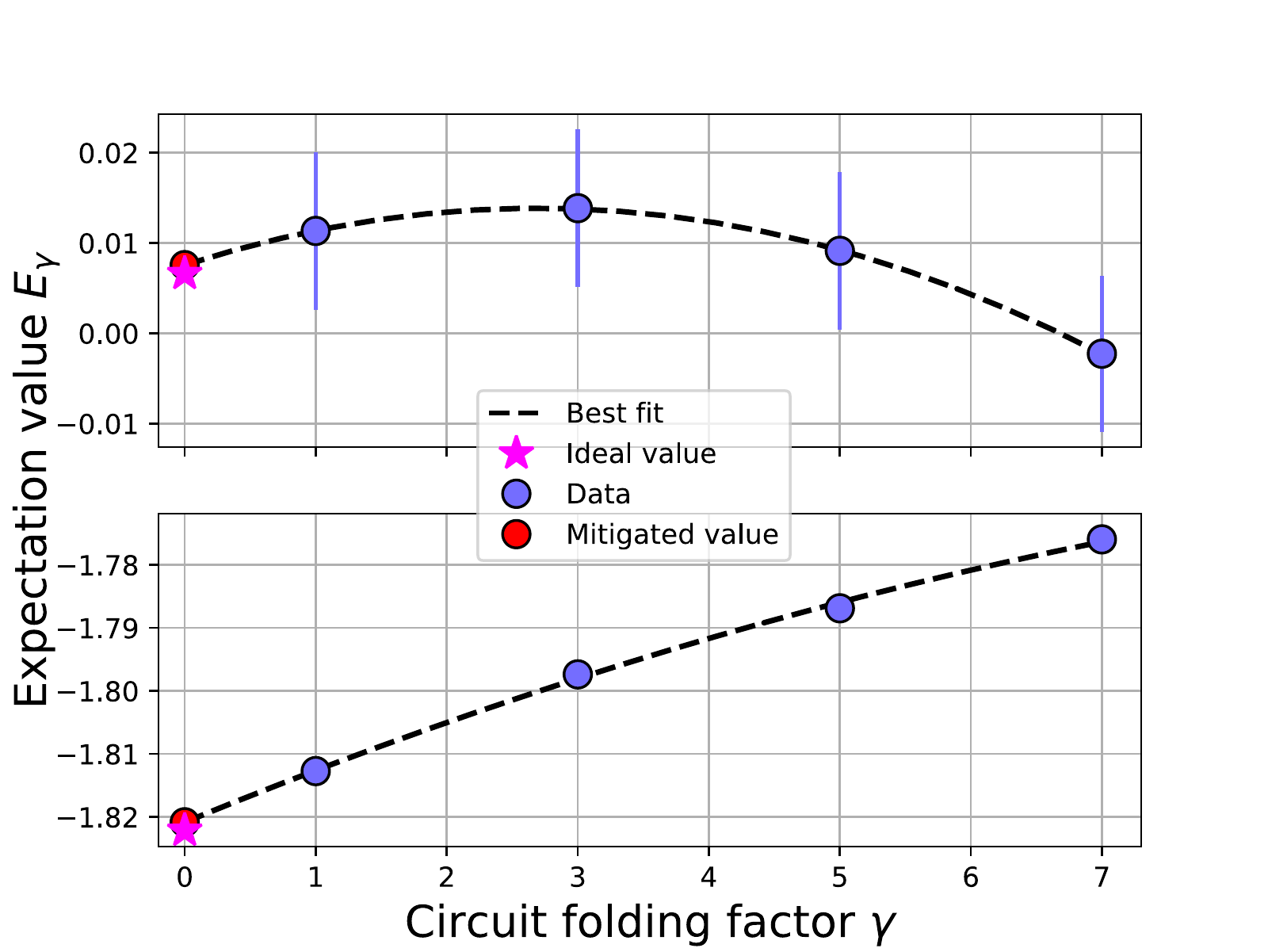}
    \caption{Example of zero noise extrapolation curves coupled with SPAM mitigation on the expectation value of two Pauli commuting groups $\mathcal{G}_2$ (top) and $\mathcal{G}_1$ (bottom). Noise scaling is performed by unitary folding with folding factors 1, 3, 5 and 7, and extrapolation is performed using a quadratic model.}
    \label{fig:ZNEfit}
\end{figure}

An example of this procedure is given in figure \ref{fig:ZNEfit}.

\section{Computational details}

\subsection{State tracking}

Both the energy and the wavefunction show continuity along the potential energy surface (PES) of the same electronic state. 
Therefore, the final energy and final angles for one molecular geometry are good starting points for another close molecular geometry along the PES. We take advantage of this property by setting $\omega$ to the previous energy computed on the PES, as well as setting the initial $\bm\theta$ parameters to the angles found in the previous calculation. In other words, with k index representing the evolution along the PES,
\begin{equation}
    \bm\theta^{\mathrm{init}}_{k+1} = \bm\theta^{\mathrm{opt}}_k
\end{equation}

sets the initial trial wavefunction for point $k+1$ as the final wavefunction of point $k$, and
\begin{equation}
    \omega_{k+1} = E_k
\end{equation}
sets the target energy for the next point to the previous energy computed.

This state tracking method allows one to reduce the optimization time. To ensure continuity of the wavefunction along the PES, it is necessary to have continuous molecular orbital (MO) coefficients as the geometry changes. The MO coefficients are pre-computed for each geometry using the PySCF package \cite{Sun2017}. Phase jumps are possible in the RHF computation, as the MO phase can freely change between independent calculations. These phase jumps do not affect the energy, but they do break the continuity in wavefunction, lowering the effectiveness of state tracking. To avoid phase jumps in the MO coefficients, we compute at each step between close geometries $k$ and $k+1$ :
\begin{equation}
    P_{k,k+1} = C_k^\dagger S_k C_{k+1}
\end{equation}

with $C_k$ the MO coefficient matrix at geometry $k$ and $S_k$ the overlap matrix at geometry $k$. The P matrix has a diagonal with $\approx\pm1$ elements. Negative signs indicate a phase jump between the two geometries. In this case, we rectify the phase of the corresponding MOs in $C_{k+1}$ matrix, and use the rectified MOs in the calculation. This ensures continuity in the ansatz parameters and allows us to facilitate convergence and reduce computation time by employing the state tracking method.

We observed that state and energy tracking, beyond reducing computation time, can also help the optimization convergence for points where it initially fails. When noticing non-converged points on the PES, one can start from a previous converged point and state track towards the desired geometry with smaller geometry steps. This technique usually allows one to obtain better convergence. However, it requires a large number of calculations, since the step size needs to be small enough to allow good continuity.

\subsubsection{Preventing jumps between electronic states}
To some extent, state tracking helps prevent jumps between close electronic states as the wavefunction tends to be continuously evolved along the PES, which is particularly useful when degenerate or quasi-degenerate states are present.
However, we observed that jumps still occurred in our computations when the $\omega$ parameter was closer to another electronic state for a particular geometry (which is particularly frequent in the presence of large energy gradients or when two electronic states are very close in energy). This behavior is expected for the FS method, but it can be undesirable when trying to follow a particular electronic state on the PES. To prevent jumps, a continuity constraint term can be added to the optimized cost function, ensuring continuity of the parameters along the PES :

\begin{equation}
\resizebox{0.88\columnwidth}{!}{$
    F(\bm\theta_k,\bm\theta_{k-1})  =  \bra{\Psi(\bm\theta_k)} (\hat{H}-\omega)^2 \ket{\Psi(\bm\theta_k)} + \eta |\bm\theta_{k-1} - \bm\theta_k|$}
\end{equation}

with $\eta$ a scaling factor that can be adjusted to balance the relative weights of the two terms in the cost function. This regularization method was used for some isolated points in the PEC, and $\eta$ was fixed to 0.1 in our implementation, based on empirical trials to find a value that influences optimization toward a region close to the previous calculated point, while still maintaining the right landscape and minima. 

\subsection{Classical optimization}\label{optimizer}

The optimization of the variational parameters $\bm\theta$ is performed using a classical optimizer. The dimensionality of the ansatz and the presence of noise in the cost function make the optimization difficult.

In addition to the finite sampling error that is inherent in quantum computing, NISQ hardware is characterised by the presence of noise within the quantum circuit, making robustness an important feature of quantum algorithms to be applied to current hardware.

Noise-tolerant optimizers are therefore the most appropriate. We use the SPSA (Simultaneous Perturbation Stochastic Approximation) optimizer for QASM (Quantum Assembly Language) simulations. In addition to being noise tolerant, SPSA also has a constant number of 2 evaluations per iteration that does not scale with the number of parameters \cite{Bhatnagar2013}.
SPSA uses a stochastic procedure to update the parameters: at each iteration, the perturbation $\Delta\bm\theta$ is a randomly generated vector that has a component in every dimension of the optimization problem. The cost function is evaluated at $(\bm\theta + \Delta\bm\theta)$ and at $(\bm\theta - \Delta\bm\theta)$, and the numerical gradient for each parameter is approximated only from these two measurements. 
Because the perturbation vector is randomly generated, additional shifts due to noise in the cost function have a minor impact on the optimization process. The noise is, in a sense, absorbed by the stochasticity of the optimizer. 

\subsubsection{Scaling of parameters}
In the UCC ansatz, the variational parameters are angles in rotation gates, ranging from $-\pi$ to $\pi$. To assist the classical optimizer in finding the optimal angles, we scale the parameters to range from $-c\pi$ to $c\pi$ with $c$ a predetermined constant. This simple procedure helps prevent optimization failures caused by optimization steps being too small to obtain non-zero numerical gradients, especially for QASM simulations.

\subsubsection{Shots scheduler} \label{subsub:scheduler}
Each evaluation of the cost function $F_\theta$ is performed through a number $s$ of state preparation and measurement procedures, where $s$ is named the number of shots. The more shots $s$ are taken, the more precise the measured estimate of the cost function. The precision follows $\epsilon \sim \frac{1}{\sqrt{s}}$ (see appendix \ref{app:precision}). \smallskip

When measurement accuracy is not crucial, it is advantageous to use a smaller number of shots, as a large $s$ requires a significant amount of computational time.
For this reason, we use an increasing number of shots during the optimization, leading to uncertain measurements far from the optimum where the cost gradients are large and increasing precision when approaching the optimal parameters. For all computations, we used an inverse exponential scheduler of the form:

\begin{equation}
s = s_{\mathrm{max}} - (s_{\mathrm{max}} - s_{\mathrm{min}}) \times e^{-k \times \mathrm{iteration}}
\end{equation}

\noindent with $k>0$ that brings the number of shots from $s_{min}$=1000 to $s_{max}$=10000 with an exponential trend as iterations are performed.

The final measurement after optimization convergence and post processing of the parameters (see section \ref{postop}) is performed with 30000 shots, allowing one to attain a better estimate of the final wavefunction and energy.

\subsection{Post optimization processing} \label{postop}

\subsubsection{Quadratic fitting}\label{subsec:quadfit}
The characteristics of the UCC ansatz search space can be harnessed to improve the parameters $\bm\theta_{\mathrm{opt}}$ found by the optimizer. 
Let us consider the energy space in the UCCSD formalism, having $\dim(\bm\theta)$ dimensions. In this representation, the parameters $\bm\theta_{\mathrm{ideal}}$ corresponding to eigenstates of the electronic Hamiltonian are located at minima or maxima in the energy space for the relevant excitations. More precisely, the eigenstates are located on vertices of parabolas in the energy space.

It is possible to take advantage of the particular location of the eigenstates to refine the solutions found by the optimizer. After optimization, one can probe the energy space around each parameter. If the solution found by the optimizer is close to an eigenstate, the energy space around each $\bm\theta$ should be a parabola. It is therefore possible to sample a few points around the optimized solution, fit a quadratic equation, and choose the vertex of the fitted parabola as a refined solution. 
We employ this method as a post-processing step to improve the cost function. The refinement is usually very low (as the optimizer already locates the vertex of parabolas well enough), but in some cases a few tenths of a percent can be gained on the cost function.

\subsubsection{Rounding of parameters}

When the electronic wavefunction is built using the UCC ansatz, it is common for some excitations to be irrelevant because the wavefunction usually does not contain all possible Slater determinants in the molecule search space. As a result, some parameters of the quantum ansatz have an optimal value of zero, and therefore do not participate in the circuit. This can be harnessed to reduce the depth of the ansatz circuit as described in Ref~\citenum{Filip2022}. Similarly, some parameters can have an ideal value of $\pi$ or $\frac{\pi}{2}$ when one determinant is completely excited to another, or two determinants have the same contribution to the wavefunction, respectively. This is particularly common when considering systems with internal symmetries.
Here, we take advantage of this feature to improve the accuracy of the optimizer solution, by including a rounding post-processing routine. After the optimization (and quadratic fitting, see \ref{subsec:quadfit}) of the ansatz parameters, we detect close to zero (or close to a fraction of $\pi$) parameters, and evaluate the cost function when rounding those angles to zero (or to the corresponding fraction of $\pi$). If the cost function is improved by rounding them, the adjusted parameters are maintained. This procedure usually only improves the result very slightly, but the refinement can go up to a few hundredths of a percent in the cost function.

\section{Results}

\begin{figure*}[ht]
        \centering
        \begin{subfigure}[c]{0.475\textwidth}
            \centering
            \includegraphics[width=\textwidth]{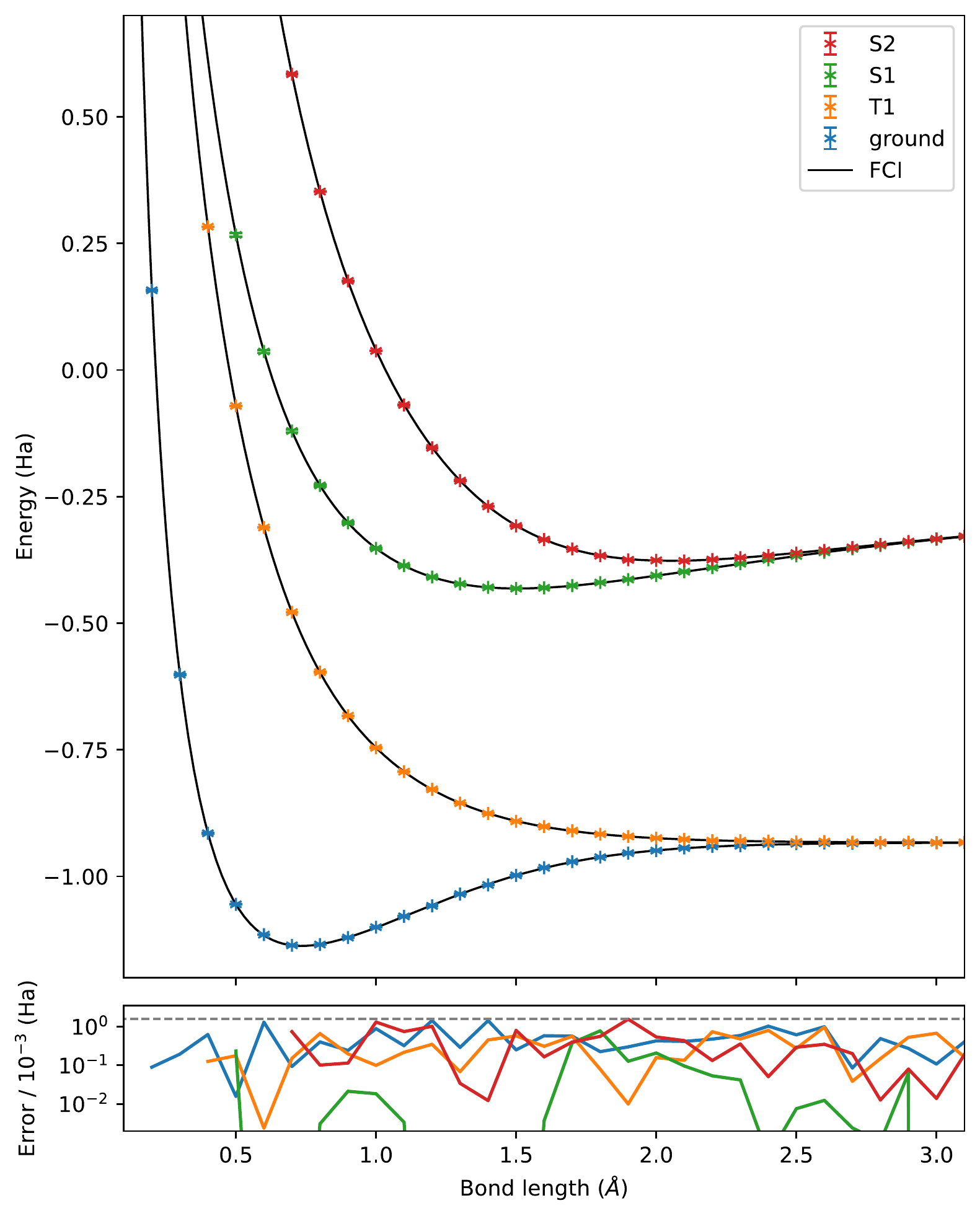}
            \caption{Potential Energy Curves of $\ce{H2}$}
            \label{fig:H2}
        \end{subfigure}
        \hspace{0.2cm}
        \begin{subfigure}[c]{0.475\textwidth}  
            \centering 
            \includegraphics[width=\textwidth]{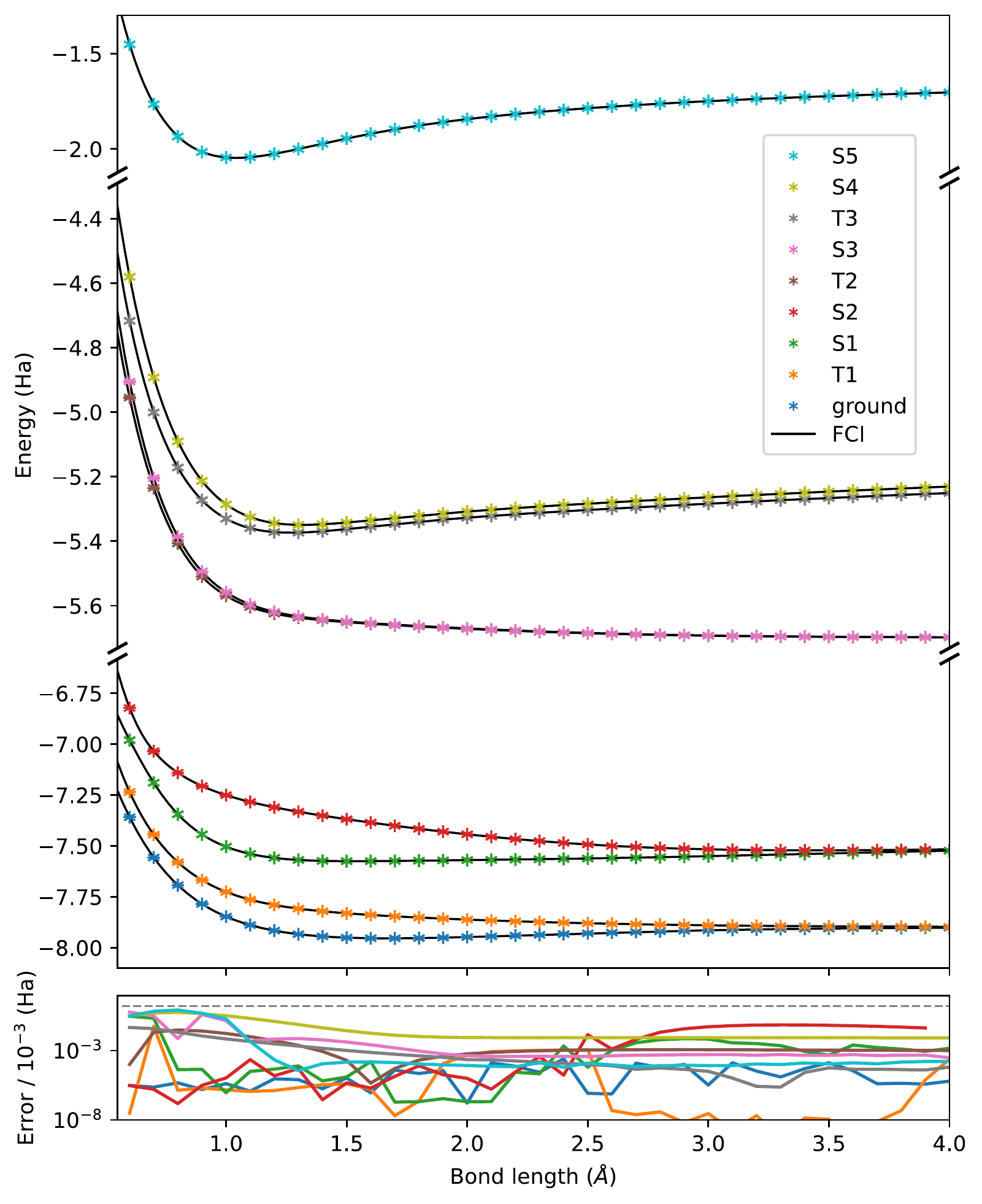}
            \caption{Potential Energy Curves of $\ce{LiH}$}
            \label{fig:LiH}
        \end{subfigure}
        \captionsetup{justification=raggedright,
        singlelinecheck=false
        }
        \caption[]
        {\small Results of FS-VQE calculations for $\ce{H2}$ (a) and $\ce{LiH}$ (b) excited electronic states. Ground-states were obtained with a standard VQE. Colored markers show the (FS)-VQE results, and solid black lines are the FCI energies obtained by numerical diagonalisation of the Hamiltonian. The lower subplots represent the \textbf{absolute error from FCI}. The error plots were rescaled for clarity, and all non-visible points are below $10^{-3}$mHa for $\ce{H2}$ and below $10^{-8}$mHa for $\ce{LiH}$.
        \footnotesize{\textbf{(a)} : Potential energy curves of $\ce{H2}$ in STO-3G basis obtained on \emph{QASM simulator} (ideal quantum computer with finite sampling \emph{shots} error) using 30000 shots for the final evaluation. The error bars of finite sampling are represented but smaller than the markers. \textbf{(b)} : Potential energy curves of $\ce{LiH}$ in a minimal s-orbitals-only basis. $\ce{LiH}$ computations were performed on a \emph{statevector simulator}, ignoring finite sampling error.} }
        \label{fig:results}
\end{figure*}

The FS-VQE method was applied on two small molecules $\ce{H2}$ and $\ce{LiH}$. The current capability of NISQ hardware (in terms of quantum volume and gate fidelity) is too limited for FS-VQE to obtain reasonable results on real quantum devices. Thus, we restricted our computations to small molecules and small active spaces that are tractable on simulators.

\subsection{Excited states of \texorpdfstring{$\ce{H2}$}{H2}}

$\ce{H2}$ was described with the STO-3G basis including the 1s orbital for each atom, resulting in 4 spin-orbitals for the system and 4 qubits after Jordan--Wigner transformation.

$\ce{H2}$ in STO-3G basis is described by two spatial orbitals $\sigma_g$ and $\sigma_u$ with up and down spin functions. The reference states we used for the 3 excited states ($T_1$, $S_1$ and $S_2$) are : 
\begin{itemize}
    \item $T_1$ : $(\sigma_g)(\sigma_u)$
    \item $S_1$ : $(\sigma_g)(\overline{\sigma_u})$ + $(\overline{\sigma_g})(\sigma_u)$
    \item $S_2$ : $(\sigma_u)(\overline{\sigma_u})$.
\end{itemize}

The UCCSD circuit for this system of 4 spin-orbitals and 2 electrons is composed of 3 excitation operators that can be implemented in a compiled quantum circuit of depth 71 (with 44 CNOT gates). Computations were performed on Qiskit's QASM simulator acting like an ideal noiseless quantum computer, including finite sampling error. A shot scheduler between 1000 and 10000 shots was used during optimization (see section \ref{subsub:scheduler}), and the final measurement was performed with 30000 shots.
Figure \ref{fig:H2} shows the results of the FS-VQE algorithm for excited states $\ce{H2}$ compared to the exact FCI states in the same basis in solid black lines. The FCI energies were obtained by numerically diagonalizing the electronic Hamiltonian matrix to obtain its eigenvalues. The ground-state results were obtained with standard VQE.
FS-VQE allows recovering the complete potential energy curves for the 3 excited states of $\ce{H2}$, at chemical accuracy. The absolute error is shown in the subplot of figure \ref{fig:H2}.

\subsection{Excited states of \texorpdfstring{$\ce{LiH}$}{LiH}}

$\ce{LiH}$ is a 4 electron system that can be described with 6 spin-orbitals in a minimal basis (considering s orbitals only for both atoms). Its excited energies were computed with FS-VQE using 6 qubits in Jordan--Wigner mapping. In this configuration, the UCCSD gate includes 8 excitations, resulting in a circuit of depth of 311 with 212 CNOT gates.
The computations were performed on Qiskit's Statevector simulator allowing the measurement of the qubits' exact state, thereby avoiding finite sampling error.
Note that the results $\ce{LiH}$ may differ greatly from the experimental data because the basis set only includes s orbitals, which is a poor approximation for the lithium atom. This minimal description allows us to put the FS-VQE to the test but does not aim for physically accurate results.

Here, $\ce{LiH}$ is described by three spatial orbitals $\sigma_1$, $\sigma_2$ and $\sigma_3$, each with up and down spin functions. The references we used for each excited state are as follows:

\begin{itemize}
    \item $T_1$ : $(\sigma_1)^2 (\overline{\sigma_2})^1 (\overline{\sigma_3})^1$
    \item $S_1$ : $(\sigma_1)^2 (\sigma_2){}^1 (\overline{\sigma_3})^1$ + $(\sigma_1)^2 (\overline{\sigma_2})^1 (\sigma_3)^1$ 
    \item $S_2$ : $(\sigma_1)^2 (\sigma_3)^2$
    \item $T_2$ : $(\overline{\sigma_1})^1 (\sigma_2)^2 (\overline{\sigma_3})^1$
    \item $S_3$ : $(\sigma_1)^1 (\sigma_2)^2 (\overline{\sigma_3})^1$ + $(\overline{\sigma_1})^1 (\sigma_2)^2 (\sigma_3)^1$
    \item $T_3$ : $(\sigma_1)^1 (\sigma_2)^1 (\sigma_3)^2$
    \item $S_4$ : $(\sigma_1)^1 (\overline{\sigma_2})^1 (\sigma_3)^2$ + $(\overline{\sigma_1})^1 (\sigma_2)^1 (\sigma_3)^2$
    \item $S_5$ : $(\sigma_2)^2 (\sigma_3)^2$.
\end{itemize}

Figure \ref{fig:LiH} shows the results of FS-VQE for the potential energy curves of $\ce{LiH}$. The solid black lines are the FCI states, obtained by diagonalization of the Hamiltonian. Absolute errors compared to FCI are presented in the subplot.

\subsection{Error mitigated results}

\begin{figure}[h]
    \centering
    \includegraphics[width=\linewidth]{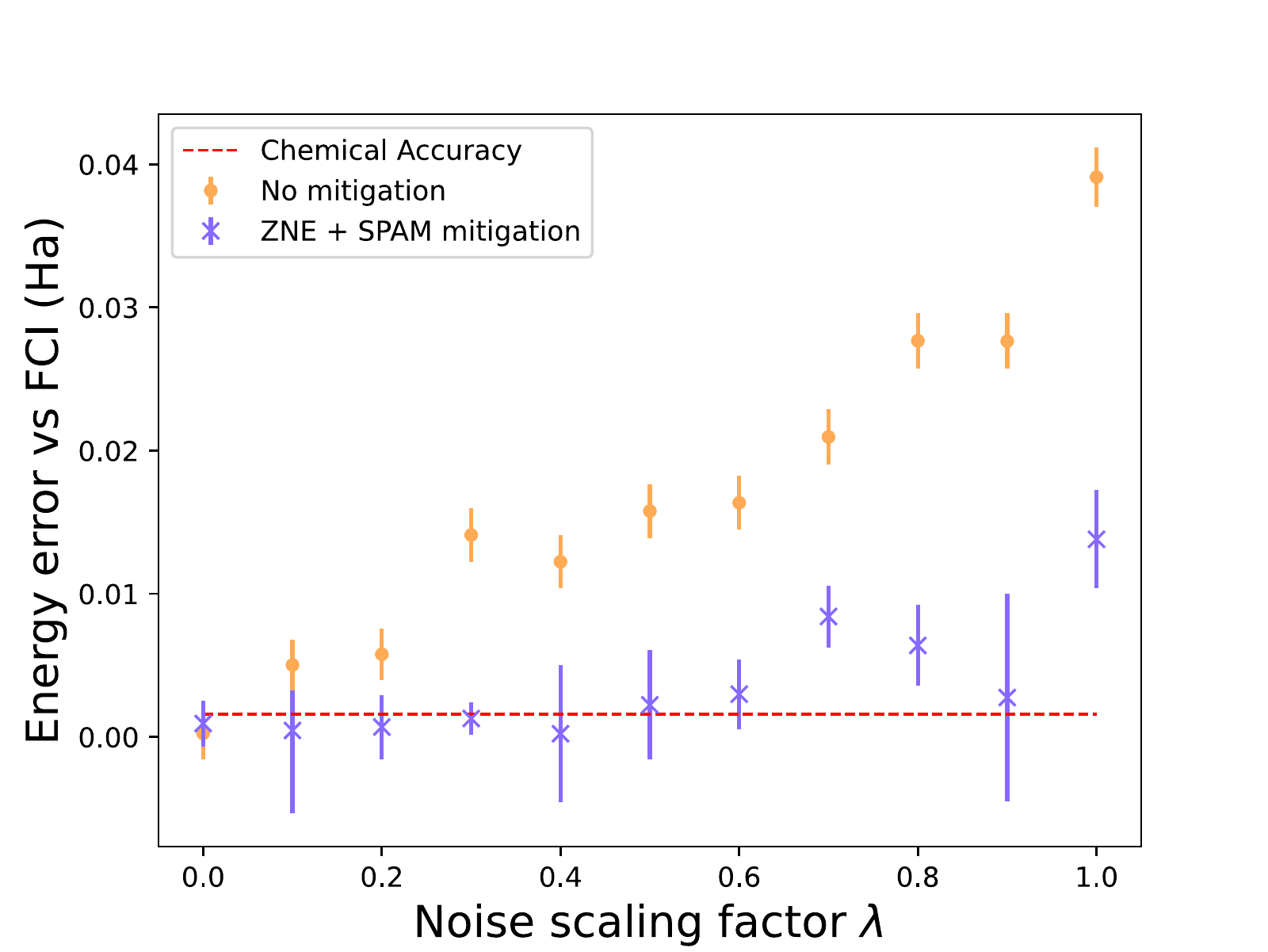}
    \caption{Comparison of the final energy accuracy obtained with non-mitigated and mitigated FS-VQE on the $S_2$ state of $\ce{H2}$ at 0.74\AA. Horizontal axis is the simulated level of noise.}
    \label{fig:scalednm}
\end{figure}

The mitigated FS-VQE algorithm was used to compute the highest excited state $S_2$ of the $\ce{H2}$ molecule in STO-3G basis, at a fixed bond length of 0.74\AA, while adjusting the level of noise with the scaling factor $\lambda$ as described in section \ref{subsec:mitig}. 20000 shots were taken for each circuit evaluation.
Figure \ref{fig:scalednm} compares the results of noisy FS-VQE simulations without mitigation with the corresponding results obtained with combined SPAM and ZNE mitigation methods.

The accuracy of the excited $S_2$ state energy computation is improved by employing mitigation techniques. When using mitigation, the energy computation reaches chemical accuracy compared to FCI for $\lambda <0.4$, while only $\lambda =0$ is below 1 kcal/mol without mitigation. The noise parameters for $\lambda=0.4$ are given in Table \ref{tab:noiseparams}.

This may indicate that error mitigation techniques will be useful tools throughout the early fault-tolerant era of QPUs to extract more accurate data from noisy quantum devices.
Our results suggest that chemical accuracy for small systems using our algorithm and the described mitigation methods could be reached for quantum devices with improved performances of about an order of magnitude compared to those of the present machines. In the long term, we expect noise mitigation methods to be one of the tools in the error correction arsenal for the early FT era, but logical qubit-based error correction techniques will be inevitably required for large-scale quantum computing.

These experiments are a proof of concept that the FS-VQE algorithm can be combined with mitigation techniques to deal with noise in quantum computations. More detailed and extensive analysis of the use of mitigation techniques in FS-VQE is left for further studies.

\section{Discussion}

\subsection{Accuracy}

In principle, the FS-VQE method with the UCCSD ansatz allows one to recover Coupled Cluster accuracy, provided that the optimization converges. For small systems such as $\ce{H2}$ or $\ce{LiH}$ with frozen core, CCSD is complete, and in theory we can recover FCI energies in the selected basis. However, for larger systems with more orbitals, single and double excitations are not generally sufficient to reach FCI accuracy and CC accuracy is expected when implementing UCCSD ansatz. Larger simulations or experiments would be needed for confirmation, as our work is restricted to very small systems.

In general, our simulations achieve good accuracy on simulated noiseless quantum computer, and all points in figure \ref{fig:results} have an error within the range of chemical accuracy (1 kcal / mol) compared to FCI, as evidenced by the subplots. However, a wide range of errors (between 1.5 mHa and $10^{-16}$ mHa) and different patterns of error curves are observed. These differences can be attributed to the optimization process: the optimization terminates when a threshold value ($10^{-9}$ in our implementation) is reached in the cost function gradient, leading to various stages of convergence between different runs. 

 When noise models are included in the simulations, the addition of mitigation techniques is required to reach chemical accuracy. A simple implementation of SPAM mitigation and zero-noise extrapolation is sufficient to significantly improve the results of FS-VQE, as evidenced in figure \ref{fig:scalednm}. This result is promising for the next early fault-tolerant quantum era with lower error rates, where error mitigation is expected to play a major role and where FS-VQE could produce useful mitigated results.

\subsection{Scaling and cost}

One asset of the FS-VQE method is that the same ansatz circuit can be used for ground-state and excited state calculations. Excited states only require the evaluation of additional Pauli operators, meaning additional state preparation and measurement (SPAM) procedures using the same hardware requirements and the same quantum circuit structures (with possibly varying post-rotation gates, representing minor changes).

The FS-VQE method has the disadvantage of involving the Hamiltonian square, making the number of Pauli string expectation values to evaluate larger than in standard VQE. After a Pauli grouping procedure (described in section \ref{subsec:pauligroup}) the resulting number of required measurements is greatly reduced, as shown in figure \ref{fig:neval}.

Many questions remain open about the feasibility of the FS-VQE algorithm (and variational quantum algorithms in general) for large systems, as the classical part of the hybrid algorithm potentially retains intractable stages for large systems. Among these may be mentioned the number of measurements needed, the classical storage of measurement results, or the pre-computation of the qubit Hamiltonian and of the FS operator. In particular, the Pauli reduction of the FS operator is a difficult classical task that would need further investigation to become scalable. The Pauli grouping procedure is also a crucial challenge, as it was shown to be a NP-hard problem \cite{Gokhale2020}, while the feasibility of the quantum FS method is highly dependent on it. In addition, optimizing variational parameters becomes more and more challenging as the system size increases, especially due to Barren plateaux. A careful design of the ansatz can overcome some of these challenges \cite{Grimsley2023}.

The scaling analysis of the proposed algorithm can be divided in terms of the number of circuit evaluations required, the circuit depth, and finally the number of shots to reach a target accuracy. The number of circuit evaluations corresponds to the number of Pauli groups, which scale as $\mathcal{O}(N^6)$ (discussed in section \ref{subsec:pauligroup}) with respect to system size.
The circuit depth depends on the ansatz. In this work, we implement the UCCSD ansatz that has a $\mathcal{O}(N^4)$ scaling in circuit depth, corresponding to the number of double excitations in the cluster operator. However, several ansätze have been proposed in the literature \cite{Magoulas2023,Burton2023} that show a more favorable scaling, such as $\mathcal{O}(N)$, while maintaining the same accuracy as UCCSD. Finally, the number of shots needed to achieve a certain accuracy $\epsilon$ scales as $\mathcal{O}(\frac{1}{\epsilon^2})$ (see Appendix \ref{app:precision}) for each Pauli string. 
Several techniques have been developed to address the measurement problem in VQAs, including Pauli grouping, measurement weighting, or shadow tomography \cite{VQEReview2021}.

The Folded Spectrum method is a general minimisation procedure that can be implemented within algorithms other than VQE to find excited states. Pauli operators are the building blocks of gate-based quantum computing, and we expect Pauli grouping procedures to remain relevant beyond variational algorithms, in which case Spectrum Folding could be one advantageous method, beyond variational algorithms, to compute molecular excited states on quantum computers. 

Employing mitigation techniques seems to be central to obtaining meaningful results on noisy devices, but it comes at the cost of running deeper circuits, with more shots. Our implementation of ZNE requires four times more shots compared to the non-mitigated algorithm, for circuits of depth multiplied by $\gamma$=1,3,5,7. This overhead is non-negligible and needs to be addressed in the future.

\section{Conclusion and prospects}

In this work, we demonstrate that Folded Spectrum method is a successful approach for computing excited states using the VQE algorithm. The concept of evaluating the FS operator instead of the Hamiltonian to reach excited states is a well-known technique in QMC, and it could also be extended to algorithms other than VQE in quantum computing. Moreover, the Folded Spectrum procedure is agnostic to the quantum ansatz and fermion-to-qubit mapping scheme, and future improvements at any stage of the VQE algorithm can directly benefit this method.

Folded Spectrum allows one to directly compute any excited state around a target energy, which is a considerable asset compared to other methods where excited states are computed sequentially. This advantage is especially important for studying larger systems that have an increasing number of electronic states. It can be particularly useful for the computation of highly excited electronic energies and of great interest for the study of photo-chemical processes and light-matter interaction. 

A major challenge to enable the scaling of variational methods is the preparation of good reference states for larger systems, which is particularly challenging for multi-reference states. As explained in Ref~\citenum{Lee2023}, a severe limitation to the scalability of ground-state VQE is the exponentially vanishing overlap of local reference states with the FCI ground-state. Our method also faces this issue, and further research is needed to allow for a systematic and scalable determination of reference states. However, we believe that finding excited states could be a less difficult task if the approach is to compute any excited state around a target energy rather than to search for a specific state (like it is the case for ground-state computations). In this scenario, our algorithm could in principle find a local minimum in the Folded Spectrum landscape that would correspond to an excited eigenstate. The increasing density of states in larger systems would benefit this approach and the reference state would thus play a less crucial role.

The main limitation of the FS method is the need to evaluate a squared Hamiltonian. In the quantum computing formalism of fermion-to-qubit mapping, this disadvantage can be alleviated by partitioning Pauli operators into commuting groups that can be evaluated simultaneously. The resulting number of evaluations needed to compute the FS operator expectation value is substantially reduced by the Pauli grouping. Despite this improvement, the number of shots required by the FS-VQE method for large systems is still prohibitive on quantum hardware, and further progress is needed to make the method scalable for practical applications. This is particularly central when dealing with noisy quantum processors, as error mitigation or error correction techniques are key to obtaining meaningful results, but come at the cost of additional quantum resources both in number of shots and number of qubits.

\section{Acknowledgement}
The authors thank Dr. David Muoz Ramo for his scientific contribution, and we also thank Dr. Maria-Andreea Filip, César Feniou, Dr. Daniel Graf, and Chiara Leadbeater for useful discussions. LCT thanks Quantinuum and École Normale Paris-Saclay for funding.

\renewcommand*{\bibfont}{\normalfont\footnotesize}
\bibliographystyle{ieeetr}
\bibliography{main}

\appendix
\setcounter{equation}{0}
\renewcommand{\theequation}{\thesubsection.\arabic{equation}}

\vfill

\subsection{Measurement precision}\label{app:precision} 

Measurement precision is directly related to the number of shots taken s. Let us consider the evaluation of $\langle \hat{O} \rangle$ with spectral decomposition :
\begin{equation}
    \hat{O} =  \sum_{i=1}^{2^n} o_i\ketbra{\Phi_i}{\Phi_i}.
\end{equation}

The measurement of $\bra{\Psi}\hat{O}\ket{\Psi}$ relies on many repetitions of preparation of and measurement to evaluate the populations $\lvert\braket{\Phi_i}{\Psi}\rvert^2$. Let V=$\langle \hat{O} \rangle$ be the target value of the process.
\bigskip

$\langle \hat{O} \rangle$ can be interpreted as the expected value of a random variable X having possible outcomes \{$o_i$\}$_{i=1}^{2^N}$, with probabilities given by the Born rule :

\begin{equation}
p(X=o_i)=\lvert\braket{\Phi_i}{\Psi}\rvert^2.
\end{equation}
Each shot of the experiment is a measure of X. By taking s shots we obtain a set of results $X_1$,...,$X_s$.
Thus,
\begin{equation}
   V = \sum_i o_i \; \lvert\braket{\Phi_i}{\Psi}\rvert^2 = \sum_i o_i \; p(X=o_i)
\end{equation}
is approximated by 
\begin{equation}
    \frac{\sum_{i=1}^s X_i}{s} .
\end{equation}

In this formalism, Chebyshev's inequality states that :

\begin{equation}
    p\left(\left|\frac{\sum_{i=1}^sX_i}{s}-V\right| \geq \epsilon\right) \leq \frac{\sigma^2}{(s-1)\epsilon^2}
\end{equation}

with $\epsilon$ the precision of the result and $\sigma^2$ the variance of X, making $\frac{\sigma^2}{s-1}$ the variance estimate of the sample by means of the central limit theorem.

$\sigma^2$ can be bounded by a constant \cite{McClean2014}, so it can be deduced that in a worst-case scenario :

\begin{equation}
    s \sim \frac{1}{\epsilon^2} \; or \; \epsilon \sim \frac{1}{\sqrt{s}} .
\end{equation}

Consequently, for a number of shots s, the precision on the expectation value $\bra{\Psi}\hat{O}\ket{\Psi}$ is of the order of $\frac{1}{\sqrt{s}}$. The extraction of classical information from a quantum system is therefore limited by a finite number of shots. This result is a direct consequence of the probabilistic nature of quantum mechanics. 
\newpage

\onecolumngrid

\subsection{Effectiveness of Pauli grouping on some examples} \label{app:pgroup}

\begin{table}[!ht]
    \centering
    \setlength{\tabcolsep}{8pt} 
    \begin{tabular}{ccccccccccc}
    \toprule
        Molecule & $e^-$ & Basis & Qubits & \multicolumn{3}{c}{$\hat{H}$} & \multicolumn{3}{c}{$(\hat{H}-\omega)^2$} \\
        \cmidrule(lr){5-7} \cmidrule(lr){8-10}
         & & & & Paulis & JW & BK & Paulis & JW & BK \\
    \midrule
        $\ce{H2}$ & 2 & STO-3G & 4 & 15 & 5 & 2 & 24 & 9 & 3 \\
        $\ce{LiH}$ & 4 & s only & 6 & 118 & 29 & 38 & 417 & 65 & 88 \\
        $\ce{LiH}$ & 4 & STO-3G & 12 & 631 & 136 & 211 & 25542 & 2216 & 3460 \\
        $\ce{BeH2}$ & 6 & s only & 8 & 193 & 43 & 46 & 1783 & 224 & 139 \\
        $\ce{BeH2}$ & 6 & STO-3G & 14 & 666 & 369 & 324 & 47171 & 8933 & 8325 \\
        $\ce{H2O}$ & 10 & STO-3G & 14 & 1578 & 837 & 746 & 111615 & 20393 & 19596 \\
    \bottomrule
    \end{tabular}
    \caption{Number of Pauli strings in the Hamiltonians and Folded Spectrum operators of several molecules in the STO-3G basis or in a minimal basis with s orbitals only, compared with the number of groups after \textbf{qubit-wise commutativity} (QWC) partitioning for Jordan--Wigner (JW) and Bravyi--Kitaev (BK) transformations.}
    \label{tab:pgroup}
\end{table}

\end{document}